\documentclass[a4paper,fleqn]{cas-sc}

\usepackage[authoryear,longnamesfirst]{natbib}

\def\tsc#1{\csdef{#1}{\textsc{\lowercase{#1}}\xspace}}
\tsc{WGM}
\tsc{QE}
\tsc{EP}
\tsc{PMS}
\tsc{BEC}
\tsc{DE}

\usepackage[authoryear,longnamesfirst]{natbib}
\usepackage{graphicx}
\usepackage{float}
\graphicspath{{./}} 
\usepackage{lscape}
\usepackage{longtable}
\usepackage{tabularx}
\usepackage{tabulary}
\usepackage{natbib}
\usepackage{calc,pifont}
\usepackage{array}
\newcolumntype{R}[1]{>{\raggedright\let\newline\\\arraybackslash\hspace{0pt}}m{#1}}
\newcolumntype{C}[1]{>{\centering\let\newline\\\arraybackslash\hspace{0pt}}m{#1}}
\newcolumntype{L}[1]{>{\raggedleft\let\newline\\\arraybackslash\hspace{0pt}}m{#1}}
\usepackage{lscape}
\usepackage[super]{nth}

\begin{document}
\let\WriteBookmarks\relax
\def\floatpagepagefraction{1}
\def\textpagefraction{.001}

\shorttitle{Exploring Key Aspects of Sea Level Rise and their Implications: An Overview}

\shortauthors{Elneel et~al.}

\title [mode = title]{Exploring Key Aspects of Sea Level Rise and their Implications: An Overview}

\author[1]{Leena Elneel}[orcid=0009-0005-6202-9081]
\cormark[1] 
\ead{lelneel@ud.ac.ae}

\credit{Conceptualization of this study, Methodology, Writing - Original draft preparation} 

\author[1]{M. Sami Zitouni}[orcid=0000-0001-7629-8702]
\ead{mzitouni@ud.ac.ae}

\author[1]{Husamuldin Mukhtar}[orcid=0000-0002-6340-4710]
\ead{hhadam@ud.ac.ae}

\author[2]{Paolo Galli}[orcid=0000-0002-6065-8192]
\ead{paolo.galli@unimib.it}

\author[1]{Hussain Al-Ahmad} [orcid=0000-0002-5781-5476]
\ead{halahmad@ud.ac.ae}

\affiliation[1]{organization={University of Dubai},
    city={Dubai},
    country={United Arab Emirates}} 
    
\affiliation[2]{organization={University of Milano-Bicocca},
    city={Milan},
    country={Italy}}

\cortext[cor1]{Corresponding author}

\begin{abstract}
Sea Level Rise (SLR) is one of the most pressing challenges of climate change and has drawn noticeable research interest over the past few decades. Factors induced by global climate change such as temperature increase, have resulted in both direct and indirect changes in sea levels at different spatial scales. A number of climatic and non-climatic events drive the change in sea level and impose risk on coastal and low-lying areas. Nevertheless, changes in sea level are not uniformly distributed globally due to a number of regional factors such as wave actions, storm surge frequencies, and tectonic land movement. The high exposure to those factors increases the vulnerability of subjected areas to SLR impacts. The impacts of events induced by climate change and SLR are reflected in biophysical, socioeconomic, and environmental aspects. Different indicator-based and model-based approaches are used to assess coastal areas' vulnerabilities, response to impacts, and implementation of adaptation and mitigation measures. Various studies were made to project future SLR impacts and evaluate implemented protection and adaptation approaches to help policymakers plan effective adaptation and mitigation measures and reduce damage. This paper provides an overview of SLR and its key elements encompassing contributing factors, impacts, and mitigation and adaptation measures with a case study focus on the Arabian Gulf. 
\end{abstract}


\maketitle

\section{Introduction}

Climate change is one of the most pressing challenges of the century. The increase of $CO_2$ emissions and Greenhouse Gases (GHG) caused a rise in temperature, which led to a range of far-reaching consequences for economies, societies, and ecosystems around the world. One of the most significant impacts of climate change is Sea Level Rise (SLR) \cite{7,133}. Over the past century, Global Mean Sea Level (GMSL) has been increasing at accelerating rates and is estimated to reach up to 98 cm by the end of the century \cite{47}. Intergovernmental Panel on Climate Change (IPCC) projected future values of global SLR to range from 0.61 - 1.10 m by the end of the century \cite{59,18} and up to 3 m by 2300 \cite{18} depending on the rate of GHG emissions. SLR is primarily caused by two factors: the thermal expansion of seawater, and the melting of ice sheets and glaciers \cite{133}. Other climatic and non-climatic events such as cyclone-driven events and earthquakes also contribute to changes in Regional/Relative Mean Sea Level (RMSL) \cite{18}.\\

The impacts of sea level rise are already persistent in many coastal areas globally. Coastal areas are of significant economic and environmental importance, and increasing human activities are placing additional stress on them \cite{80}. 
Low-lying coastal areas, such as Small Island Developing States (SIDS), are particularly vulnerable to inundation, storm surges, and floods \cite{18,IPCC96}. Areas exposed to tidal events also experience impacts of SLR \cite{15}. In some areas, saltwater intrusion into groundwater and freshwater aquifers threatens freshwater supplies and agricultural crop production \cite{23,18}. Furthermore, SLR can cause physical changes to the coastal areas such as coastal erosion and flooding leading to loss of land and infrastructure \cite{IPCC96,132}. To address challenges caused by climate change and SLR and reduce their impacts, a comprehensive and coordinated response is needed. This includes reducing GHG emissions to mitigate the anticipated damage of future SLR scenarios and associated extreme weather events \cite{50}, as well as investing in adaptation measures to help vulnerable communities cope with the impacts that are already occurring \cite{70} including hard or soft engineering solutions, ecosystem-based structures, and strategy plans \cite{24}. The use of visualization tools that represent the relationship between impacts and contributing factors to SLR can bridge the gap in understanding the relation between different components of SLR, raising awareness, and helping decision-makers form proper mitigation and adaptation measures \cite{80}. \\

\begin{figure}
\begin{center}
  \includegraphics [width=0.8\textwidth]{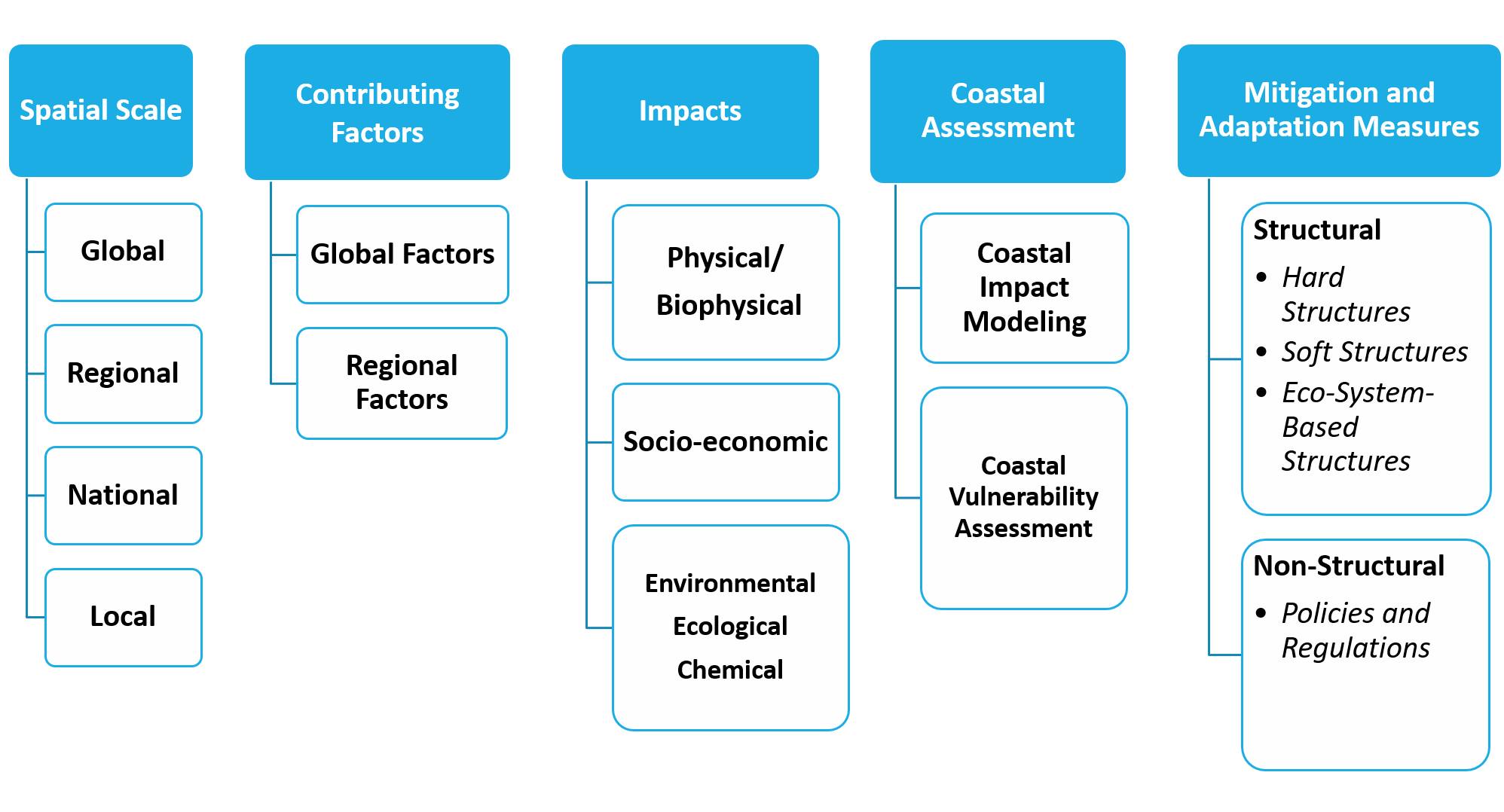}
  \caption{Classification of SLR aspects as identified and discussed in this paper}
  \label{fig: Research Diagram}
\end{center}
\end{figure}

\begin{figure}[ht]
\begin{center}
  \includegraphics [width=0.8\textwidth]{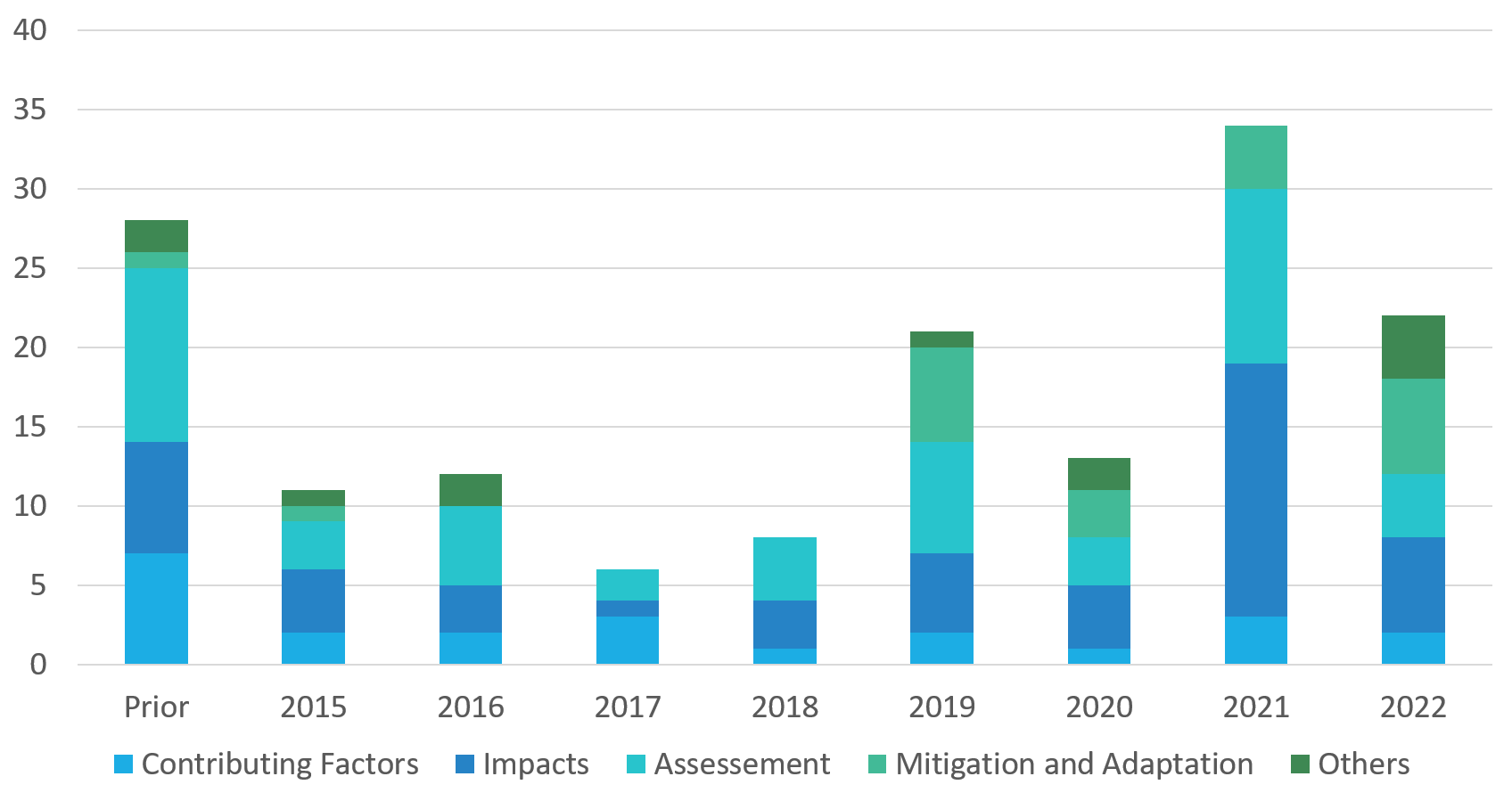}
  \caption{Number of reviewed papers of each category over the years}
  \label{fig: papers stats}
\end{center}
\end{figure}

This work presents an overview of SLR and its key elements as found in the literature. Those elements were identified based on our review of IPCC reports \cite{IPCC2015, IPCC96}. It also highlights the climatic and non-climatic drivers of SLR and its consequences. The studies in this work cover all spatial scales (global, regional, national, and local) and emphasize the distinction between global and regional contributing factors to SLR. As we found a complex interconnection relation among SLR impacts, we have grouped them into 3 categories: physical/biophysical, socioeconomic, and environmental/ecological/chemical impacts. Furthermore, due to the lack of a standardized way of categorizing different analysis approaches in the literature and the vast variations among research studies, we have identified two main categories for coastal analysis and assessment approaches, namely: coastal impact modeling approaches and coastal vulnerability assessment approaches. Figure \ref{fig: Research Diagram} shows the main aspects related to SLR that were covered in this paper. The reviewed papers were collected from different databases and publications repositories including IEEE Xplore, Science Direct, JSTOR, Springer, Elsevier, MDPI, Wiley Online Library, Francis and Taylor, and Frontier for the years between 2015 and 2022 (with the exception of papers that discuss general concepts, historical information, and highly relevant info) using the following keywords and their combination: "sea level rise", \emph{"coastal vulnerability assessment"}, \emph{"factors of sea-level rise"}, \emph{"climate change impacts on sea level rise"}, \emph{"visualizing sea level changes"}, \emph{"sea level rise impacts"}, \emph{"adaptation"}, \emph{"modeling sea-level rise"}. A screening process of the search results was conducted and over 100 references were selected for the review process with emphasis on studies that include highly vulnerable coastal areas/communities such as Southeast Asia and Mediterranean Basin countries. Papers with sole focus on SLR impacts on ecosystems such as mangroves were excluded from the review process. Nevertheless, they were briefly mentioned in sections \ref{subsec: Environmental Impacts} and \ref{subsec: Structural Mitigation Measures}. Similarly, papers that focus solely on flooding were excluded to differentiate flooding caused by SLR and climatic events in coastal regions from flooding in inland areas and riversides. Figure \ref{fig: papers stats} shows the number of papers that were reviewed in each category over the years.

%

\section{Contributing Factors}\label{sec: factors}

SLR is a significant indicator of climate change as the increasing global temperature contributes to the increase in GMSL. Thermal expansion of ocean water and glacier and ice melts were identified as the main contributors to global SLR \cite{93}. However, changes in sea levels are neither linear nor uniform \cite{116,50,133}. The propagation of changes in GMSL depends on the bathymetry of seas and oceans, which can take anywhere from less than a decade to hundreds of years to reach deep ocean regions \cite{133}. Uncertainties associated with changes in sea levels should be accounted for in SLR analysis and projections, as it was found that earlier IPCC reports did not give them sufficient emphasis. Associated uncertainties and the variation of different contributing factors can lead to changes in RMSL different from GMSL \cite{25,94,116}. This section highlights the contributing factors of various spatial scales and the associated uncertainties.

\subsection{Global Factors}\label{subsec: global factors}

\begin{figure}[ht]
\begin{center}
  \includegraphics [width=1\textwidth]{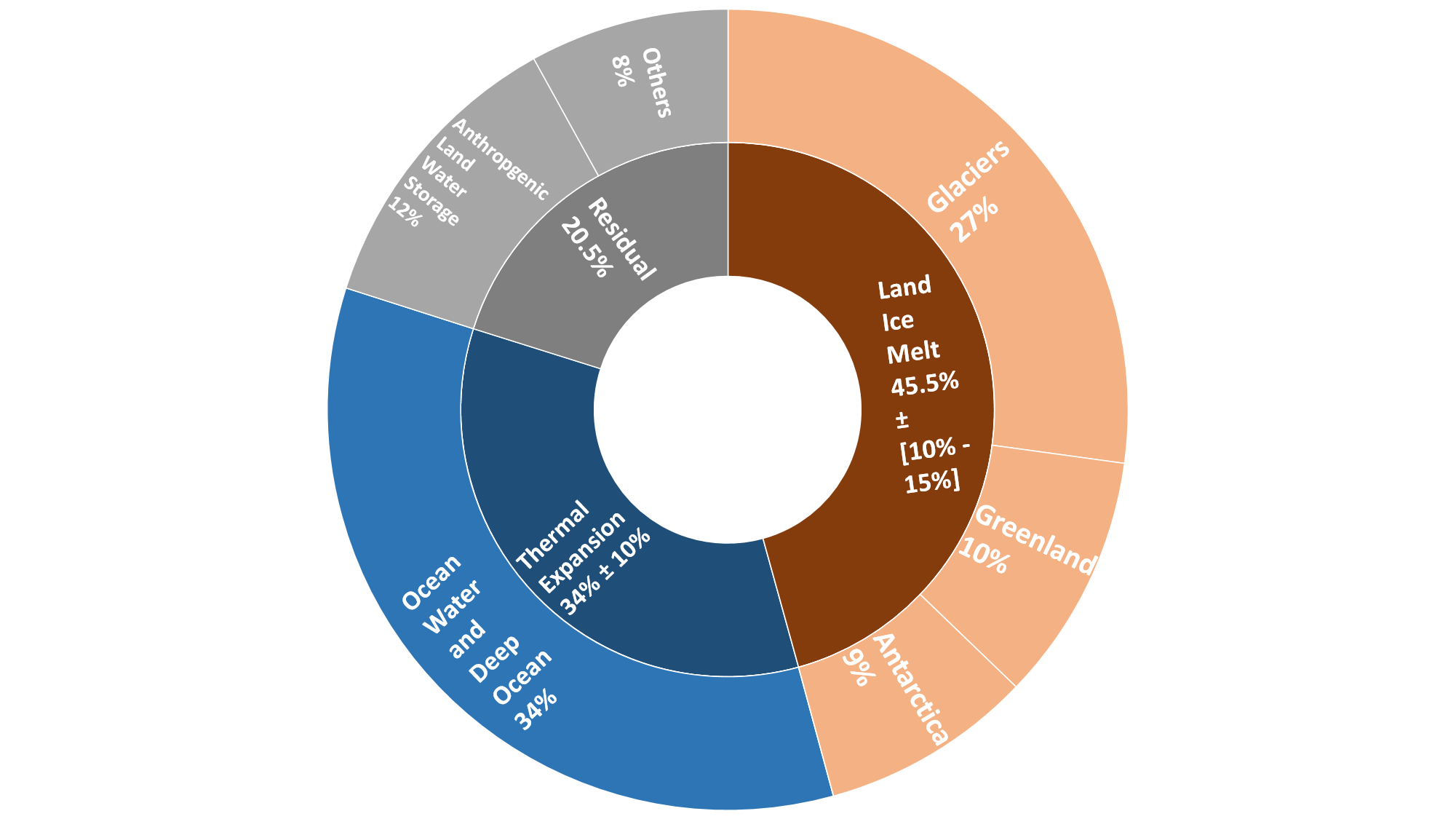}
  \caption{Factors contributing to GMSL}
  \label{fig: Contributing factors to GMSL}
\end{center}
\end{figure}

Global changes in sea level can be a result of both non-climatic (geological) and climatic changes with the latter having more influence on SLR that is noticeable \cite{7}. In 1983, Hoffman identified five climatic change factors that contribute to changes in sea level and used them to build 90 different scenarios. Hoffman factors include 1) Carbon dioxide emission, 2) Fraction airborne, which considers the movement and chemical actions of carbon between atmosphere, biosphere, and hydrosphere, 3) Concentration of other GHG specifically methane, nitrous oxide, and chlorofluorocarbons, 4) Climate sensitivity that measures the change in temperature due to gases emission, 5) Thermal expansion of ocean water that measures the transport of heat into deeper ocean water, 6) Snow and ice contribution \cite{7}. The contributing factors identified by IPCC Assessment Reports (AR) were narrowed to 1) Thermal expansion of ocean water, and 2) Ice and glacier melting. Nevertheless, \cite{58} and \cite{IPCC96} considered uncertainty associated with the factors contributing to GMSL in their projections, such as uncertainty in the future estimation of GHG. In addition to the factors listed by IPCC, \cite{116} have also identified additional factors that contribute to GMSL, including 1) Land-based ice melting (that will eventually end up in the sea), 2) Changes in land water storage due to natural and human activities, such as building dams and extracting water for agricultural activities. Those activities can result in both increase or decrease in SLR rates. Although there were no historical observational data to identify the contribution of land water storage, \cite{116} stated that it can cause a change of up to 2 mm/y during ENSO events \cite{elnino}. Furthermore, uncertainties associated with global contributing factors were considered for SLR projection. Those uncertainties include deep ocean warming (although it has a relatively small contribution to SLR compared to upper ocean warming), and dynamical instabilities in melting glaciers and ice sheets that are below sea water level, which causes an accelerated increase in SLR due to thermal expansion of ocean water. Overall, it was found that considering uncertainties of thermal expansion and melting ice sheet will cause a deviation by 10\% and 15\%, respectively, to GMSL by the year 2100, making glaciers and ice sheets the main contributors to GMSL \cite{116}. Figure \ref{fig: Contributing factors to GMSL} shows the  \cite{116} interpretation of the contributing factors to GMSL in IPCC AR5. The percentage of contributions of ice melt and thermal expansion to GMSL are 45.5\% and 34\%, respectively, where glaciers' melts form the highest share of land ice melting contribution. The estimated uncertainties associated with land ice melt and thermal expansion factors are 15\% and 10\%. The uncertainty is determined by how much the individual time series deviates from the mean. This chart also considers the contribution of the net impact of anthropogenic land water storage to GMSL and the small contribution of deep ocean water to thermal expansion. The uncertainty associated with the residual 20\% and the other contributors to GMSL were not mentioned \cite{116}. The contribution of different factors to GMSl is continuously analyzed and corrected via sea level budget analysis where observations are compared with estimated values of a sum of different factors \cite{dieng2017new}. Equation \ref{eq:sea level budget} is an example of considering different contributing factors where $GMSL_{Steric}$ refers to changes in GMSL due to thermal expansion and salinity and $GMSL_{Barystatic}$ refers to changes in GMSL due to water exchange between land and ocean such as ice melts and land water storage. Updated versions of the sea level budget can include sea level change in deep ocean due to thermal expansion \cite{wang2022} and atmospheric water vapor \cite{dieng2017new}.\\
\begin{equation}\label{eq:sea level budget}
GMSL_{Sterodynamic}= GMSL_{Steric}+ GMSL_{Barystatic}
\end{equation}

\subsection{Regional Factors}\label{subsec: regional factors}
\begin{figure}[h]
\begin{center}
  \includegraphics [width=0.8\textwidth]{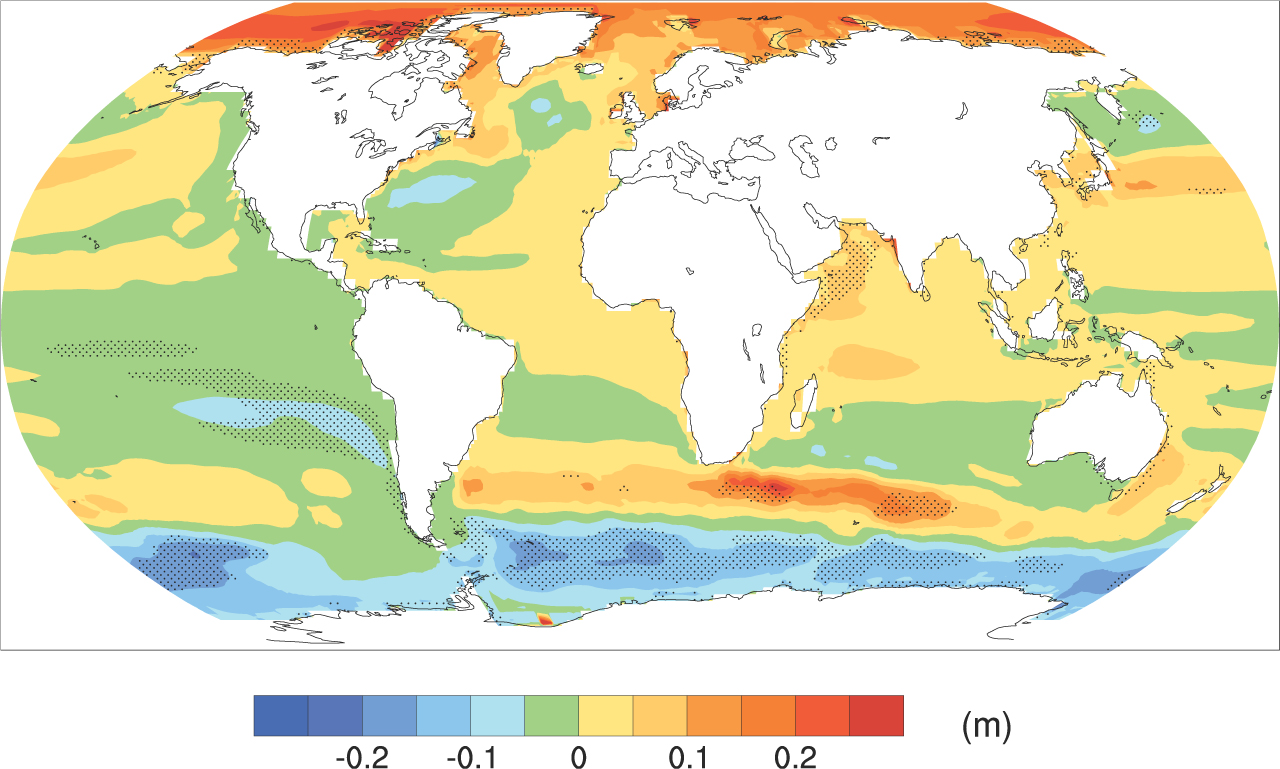}
  \caption{Variation between the average RMSL and GMSL. Positive values indicate higher deviation from the GMSL \cite{58}}
  \label{fig: Variation between GMSL and RSL}
\end{center}
\end{figure}

RMSL diverges from GMSL at 80\% of coastal areas globally \cite{carson}. The divergence between GMSL and RMSL is attributed to some adjustments in global contributing factors. Those adjustments include contributing factors that affect SLR at the regional scales and beyond such as vertical land movement, melting land-based ice, and dynamic ocean movements \cite{95}. The IPCC AR6 has considered the contribution of other factors and uncertainties that affect both GMSL and RMSL. This report lists climate drivers that impact coastal systems such as $CO_2$ concentration, Sea Surface Temperature (SST), wave actions, and rainfall run-offs \cite{IPCC96}. Overall, regional SLR is a sum of a set of climatic and non-climatic factors alongside GMSL \cite{72}. Figure \ref{fig: Variation between GMSL and RSL} illustrates the variations between RMSL and GMSL where positive values show high variation rates as can be observed near Greenland and the majority of global coastal zones (especially Southeast Asia, East Africa, and Northern Europe). However, this figure excluded Semi-Closed Seas such as the Mediterranean Sea in which numerous studies indicated that they will be highly impacted.   \\

\subsubsection{Land Movement}\label{subsubsec: RF-LM}
Land movements (up/down) due to non-climatic factors, such as tectonic movements and gravitational effects were found to influence regional SLR \cite{15,65}. Tectonic movement along with the Glacial-hydro-Isostatic Adjustments will lead to an increase in RMSL levels beyond the global rates \cite{39}. Examples of the effects of land movements can be observed in Finland (uplifting tectonic movements) and Japan (post-earthquakes effect), which have caused variations in RMSL rates compared to GMSL \cite{58}. Moreover, ground motions influence tide gauge measurements and SLR rate estimations \cite{65}. Tide gauge data are used to measure sea level with respect to the ground and can be used in monitoring land movements and regional SLR \cite{65}. Historical evidence shows that non-climatic factors have contributed to changes in RMSL and induced other extreme weather events associated with SLR. Historical evidence in the city of Venice in Italy showed that vertical land movement caused by human activities could significantly contribute to the RMSL rise, further worsening the hazard posed by SLR and extreme weather events caused by climate change \cite{127}. In Greece, massive destructive tsunami waves that impacted the coastal areas occurred post-earthquake on Crete island \cite{35}. It is also predicted that coastal areas in the Arabian Gulf will be affected by post-earthquake tsunami waves. In case a significant earthquake occurs in Markan Subduction Zone (MSZ), tsunami waves will pass through the strait of Hormuz and cause an increase in sea levels by 1 - 2 m \cite{17}. \\

\subsubsection{Dynamic Ocean Movement (Ocean Circulations)}\label{subsubsec: RF-DSL}
SLR rates are affected by the intensity, height, and frequency of wave actions in the coastal and tidally influenced areas \cite{129,130}. Whereas linear SLR scenarios and extreme events can also influence wave actions in coastal areas \cite{129}. Short-term impacts of tide-driven and storm-driven events also exhibit prominent seasonal patterns that are used to assess the changes in regional sea levels \cite{15}. Wave intensity can be calculated via Dyer function of wave power that uses components such as wave height, gravity, and density \cite{111}. Wave actions can be modeled using historical data \cite{73} and numerical simulators such as SWAN \cite{75,54,129,77}, TOMAWAC \cite{129}, XBeach \cite{130}, Method of Splitting Tsunami (MOST) \cite{35} or forecasting models \cite{129,77}. Google Earth engines and satellite images were used to monitor the dynamics of seawater via an open-source Python-based tool PyGEE-SWToolbox \cite{85}.  \\

\subsubsection{Ice Melting}\label{subsubsec: RF-LBI}
The land-based ice melting continues to contribute to RMSL in two forms: 1) The transfer of freshwater into the ocean alters the density structure of the ocean and the wave actions which results in changes in RMSL over time 2) The ocean basin deformation as an elastic response to the water mass movement from land to the ocean \cite{65}. Increasing evidence suggests that the recent negative mass balance of ice sheets is caused by the accelerated flow of outlet glaciers along certain margins of Greenland and West Antarctica and the increasing rates of discharge by icebergs into the surrounding oceans \cite{65}. The mass loss in Greenland and Antarctica is subject to Surface Mass Balance (SMB). Under the influence of dynamic discharges and ocean warming, a huge mass loss of ice sheets will occur \cite{25}. Once a certain threshold is exceeded, the interplay between decreasing surface elevation and escalating surface melting can trigger further ice loss, potentially resulting in the total loss of the Greenland ice sheet \cite{25}. Although there are uncertainties associated with future ice melting levels, they should be considered in SLR projection to increase the results' reliability. For instance, the uncertainties associated with ice melting were included in a study by \cite{39} to project SLR scenarios in the Italian peninsula that have shown higher rates of projected RMSL than scenarios that do not consider them. \\

\subsubsection{Other Factors}\label{subsub sec: RF-others}
In addition to the previous regional factors that were listed by \cite{15}, other factors that play a role in changing RMSL were found. \cite{4} identified time and distance dimension as a source of deviation between RMSL and GMSL as changes will rely on both of those dimensions to propagate globally. Since events contributing to SLR at both global and local scales are interconnected, changes can propagate through different water levels to end up reflecting in coastal processes. \cite{132} listed other climatic and non-climatic components that cover environmental, physical, and socioeconomic aspects of SLR and its impacts, including CO2 emissions, SST, characteristics of extreme weather events and rainfall, land movement and subsidence due to natural and human activities, flooding, wetland loss, impeded drainage, and saltwater intrusion. The salinity of the nearby sea also alters SLR \cite{116}. Moreover, some studies showed that there is a linear relationship between estuarine hydrodynamics and SLR \cite{110}. Ocean dynamics can potentially affect the salinity of seawater \cite{50}. Adding salinity as a factor in future SLR projections will give more accurate and realistic estimates of SLR impacts than what is estimated by scenarios that don't include such uncertainties \cite{39}. Salinity is attributed as a contributing factor alongside thermal expansion contribution in sea level budget calculations despite the fact that the changes caused by this factor become evident over longer time periods \cite{dieng2017new}. \\
SST, associated with the increasing global air temperature and thermal expansion of the oceans, increases SLR rates. For example, \cite{17} changes of SST in the Arabian Gulf as an associated factor to climate change which caused thermal expansion of seawater. The study used various data to create spatial distribution over 16 years and all seasons of the region. The results have identified three thermal zones and confirmed that the Gulf region is the warmest sea in the world (triple the global average rates). The increase of SST in what is known as the Indian Ocean Dipole was also found to increase SLR rates \cite{72}. The warming effects of ENSO (El Nino oscillations) events were also found alongside other climatic components to affect SLR in southeast Asia \cite{22,27}. \\

Another uncertainty that should be considered in projecting SLR is rainfall rates as they have a direct influence on flooding \cite{39}. SLR can also be impacted by the type of event causing it, whether it is short-term or long-term \cite{47}. For example, extreme weather events will have an impact on the wave height and intensity which can result in a short-term SLR in the form of flooding \cite{59}. Wave actions extents cause erosion on coasts, cliffs, and marine \cite{54}. Cites that are subject to be impacted by SLR and extreme weather events respond differently to those forcing factors based on a number of variations \cite{15}. A study on the contributing factors to SLR in mangrove areas found that they are highly exposed to climatic components and listed factors like SLR, increased wave, wind, increased air and water temperatures, increased rainfall activities, and freshwater availability that affect salinity processes, as the main causes that influence inundation periods and accretion rates \cite{100}. Other factors that determine the degree of impact on a mangrove area in correspondence to those climatic components were used in vulnerability assessment studies (see section \ref{subsec: Environmental Impacts}). Generally, the increasing regional SLR trends will increase the vulnerability of the exposed areas \cite{100}. \\

Other non-climatic components can also contribute to SLR rates and their impacts. The elastic response to mass loss due to land-based ice melting and water exchange between land and sea influence SLR and its impact \cite{65,116}. Moreover, human activities have contributed to physical shoreline changes and their susceptibility to SLR \cite{48, 21} in which including building dams and extracting water for agricultural purposes also play a role in changes in SLR \cite{116}. In Thailand, groundwater increased SLR \cite{58}. The increasing human activities in the oil industry have increased land subsidence that will lead to an increase in seawater levels in the low-lying coastal areas \cite{17}. Although non-climatic attributes were not a focal point in SLR studies, they have a considerable influence on SLR and future projections \cite{65}.

\section{Impacts}\label{sec: impacts}

\subsection{Physical Impacts}\label{subsec: Physical Impacts}
The physical impacts of SLR are highly visible in coastal areas where they affect physical communities directly. The physical changes to coastal areas due to SLR and climatic events occur in both short- and long- terms, where they can be irreversible in the absence of adaptation measures (see section \ref{sec: mam} for more details). Short-term physical impacts, including flooding and inundation usually occur due to extreme weather events and can cause severe damage to the impacted areas. Long-term physical impacts include coastal erosion, wetland loss, saltwater intrusion into groundwater and raising water tables, and impeded drainage \cite{58}. \\

Under extreme GHG emission scenarios, coastal areas are at risk of high erosion rates. In Krk island in the Mediterranean Sea, the increasing levels of waves' height and energy due to climatic components have caused both beach and cliff erosions \cite{54}. A higher deviation of RMSL over GMSL at Victoria Beach in Spain caused a drastic increase in erosion rates leading to an additional need for beach maintenance activities that are costly \cite{104}. Erosion caused by increased and intense waves' activities has also impacted the coastal areas at Beaufort Sea in the Arctic Ocean \cite{119}, UAE \cite{59,13}, Mostaganem shoreline in Algeria \cite{120}, Hawaiian Islands \cite{121}, India \cite{81}, Italy and shorelines in the Mediterranean sea \cite{46}, and Estonia \cite{123}. In addition to wave activities, erosion can also be caused by sediments transport from rivers and onshore overwash activities \cite{104}. Erosion rates can be predicted using Brunn Rule \cite{104,47} which is used to predict geographical changes of sandy coastlines under mean SLR situation. According to Brunn Rule, under SLR effect the coast should naturally retreat in an upward horizontal direction, such that the eroded sediment from the upper beach is deposited in the nearshore area and the resulting elevation in the bottom of the nearshore area is equivalent to SLR. Furthermore, physical impacts of SLR include wetland loss \cite{104, 47}. Although coastal wetland forms a natural protection structure for coastal areas against inundation and flooding, the increasing rates of climate-induced activities, such as SLR and wind-driven waves, makes them subject to coastal erosion and inundation which threaten the ecosystem environment they reserve \cite{86,26}. \\

Another long-term physical impact of SLR is altering coastal watersheds and groundwater discharge which results in saltwater intrusion into groundwater \cite{47}. This phenomenon is also known as water salinization. Saltwater intrusion into groundwater has been reported in Wadi Ham in UAE, due to misuse of water in agricultural activities \cite{63}. The intrusion of saltwater into groundwater was also reported in coastal areas along Gulf of Oman and Arabian Gulf \cite{61}, SIDS \cite{92}, Nile Delta in Egypt \cite{125}, and is expected to increase in West African countries in the future \cite{82}. It is worth noting that groundwater extraction approaches and other factors also contribute to the salinization phenomena. Salinity was also found to increase with the associated changes of estuarine hydrodynamic under SLR scenarios in Eastern China \cite{110}. \\

Flooding caused by SLR can lead to a rapid increase in water salinization \cite{18}. Additionally, short-term physical impacts of SLR can be observed in the form of changing tidal hydrodynamics as observed in increasing intensity and height of waves and inundation \cite{47}. SLR and extreme weather events also lead to temporary flooding and permanent flooding in the form of inundation \cite{14}. SLR and other regional factors increase the susceptibility of coastal flooding at coastal areas in the Mediterranean Sea \cite{39,46,54,20,127,128, 104}, U.S. coasts \cite{50,53}, UAE \cite{52}, Bangladesh \cite{75}, India \cite{73}, Malaysia \cite{64}. \\

Physical shoreline changes can be measured using different remote sensing data and geophysical location systems \cite{13,66,59,83,64}. Moreover, soft computing techniques can be used on remote sensing data in shoreline mapping to estimate Land Use (LU) and Land Cover (LC) which are terms used to describe the use of land for human activities and the physical characteristics of the land surface. Dwarakish and Nithyapriya \cite{83} conducted a comparison study between conventional and soft computing methods of inundation mapping using remote sensing data and concluded that Support Vector Machine (SVM) \cite{svm} algorithms show promising applications due to their adaptability and learning capacity with limited data. Data that was used to analyze shoreline changes are geomorphology, LU/LC, Elevation data, and tide gauges.\\

\begin{figure}[h]
\begin{center}
  \includegraphics [width=1\textwidth]{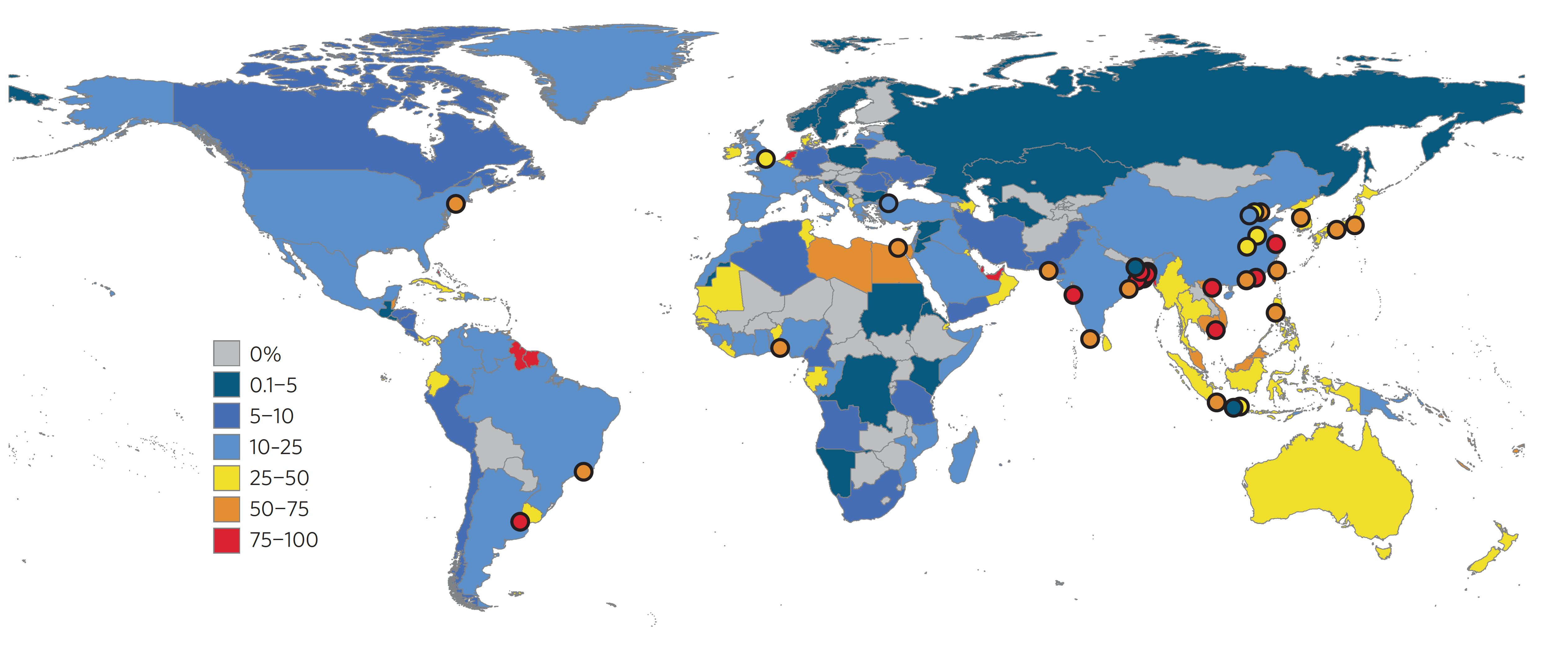}
  \caption{Areas Impacted by SLR and the affected percentage of the population \cite{133}}
   \label{fig: impacted areas worldwide}
\end{center}
\end{figure}
 \subsection{Socioeconomic Impacts}\label{subsec: Socioeconomic Impacts}
The impacts of climate change-induced SLR can vary according to the nature of the impacted areas. Low-lying lands will be highly impacted by SLR with an increasing sociological and economic risk due to the high density of population in those areas \cite{9}. In SIDS, a group of countries that consists of small islands with similar topographical nature, climate change, and SLR will result in changes in biodiversity and coastal resources which will be reflected in economic losses and negative population growth. For example, SLR in the Kingdom of Bahrain will result in a major hit on socioeconomic due to the lack of capacity and space to accommodate the changes \cite{26}. In many developed countries, the LU of the coastal areas was restructured for economic purposes to the extent of building artificial lands, such as Palm Islands in Dubai, however, these developed coastal areas are considered more vulnerable to SLR due to their socioeconomic importance. For instance, erosion caused by SLR affected properties and street infrastructure along the coast in some emirates in UAE \cite{59}. The increasing economic pressure on coastal areas makes changes in seawater levels and extreme weather events caused by climate change with significant impact on different sectors, such as transportation or even leisure activities, namely scuba diving and commercial trade \cite{112}. An approximate of 100 airports globally, that were built in low-lying lands, will be impacted by SLR under a global temperature increase scenario of $2^{\circ}C$ degrees. This impact will highly affect airports located in southeast Asia that are highly susceptible to climatic SLR. Aside from transportation, airports are viable for many sectors such as healthcare and food security which means that the impacts of SLR on airports might lead to economic paralysis \cite{109}. In the middle east, 24 coastal ports will be impacted under climate changes induced SLR and extreme weather event scenarios \cite{40}. SLR will also impact the military's marine operations and equipment installations \cite{103}. Therefore, loss of land due to the physical impacts of SLR will cause migration of coastal habitations \cite{IPCC96}. In Indonesia, it was estimated that there will be a huge impact of SLR on the economy as an increase of 100 cm of seawater will cause a land loss of 34 km$\textsuperscript{2}$. This loss of land will impact coastal building and infrastructure, and the high population density living near the coastal areas, which will result in huge national socioeconomic loss \cite{124}. In Mombasa - Kenya, SLR caused direct and indirect impacts on coastal tourism activities and Geographic Information System (GIS) based analysis showed high exposure of population and economic assets to climate variability causing a loss of billions of dollars and negative growth and distribution of population \cite{131}. Coastal areas in Some of the Gulf Cooperation Council (GCC) countries have changed in the past few years due to the construction of artificial lands for economic purposes \cite{17}. SLR will highly affect those man-built lands leading to huge economic losses. Figure \ref{fig: impacted areas worldwide} illustrates the affected areas by SLR and the percentage of the population that are at risk. \\

\subsection{Environmental, Ecological, and Chemical Impacts}\label{subsec: Environmental Impacts}
SLR and extreme climatic events will have a high impact on environmental ecosystems. The impacts on ecosystems can be perceived as both positive and negative. Generally, some ecosystems have the ability to adapt naturally to increases in SLR up to a certain rate. However, variations between the rates of SLR and the ecosystem's adaptability to cope with those changes will lead to a negative impact \cite{47}.
Marine habitats are subject to extinction due to the effect of rising SST and thermal expansion of the oceans \cite{IPCC96,17}. The increase in sea temperature due to climatic events causes coral bleaching which damages marine habitats and the biodiversity of the coral species \cite{26,62}. Changes in climatic components associated with regional SLR, such as El-Nino oscillation, led to coral bleaching and decreasing in mangrove-covered areas in the Arabian Gulf \cite{17,62}. Additionally, the increase in salinity, increase in temperature, and the decrease in oxygen levels in the Arabian Gulf have impacted the biodiversity of fish species. On the contrary, the increase in the surface and deep ocean temperature in The Arabian Gulf and Gulf of Oman caused Jellyfish outbreaks \cite{62}. In California, similar variations between positive and negative impacts on the number of fish species were noticed in correlation with the increasing SST \cite{137}. Marine ecosystems in the South China Sea are experiencing both negative and positive impacts with the regional variation of SLR \cite{55}. Coastal wetland areas are expected to decrease by 22\% globally before the end of the century, which will directly impact the ecosystem and the natural habitats they reserve \cite{26}. Chemical changes in carbon dynamics due to water intrusion will lead to metal toxicity of the soil, as well as affect the quality and volume of the agricultural crops \cite{18}. In Portugal, ecologists are concerned with water pollution in case coastal erosion in the city of Maceda paved the way for seawater to reach the nearby landfill \cite{70}. Additionally, water pollution is a potential aftermath scenario to flooding events where drinking water intersects with sewage water, causing serious health problems to human life \cite{23}. Pollution in the sea and ocean water can also occur as a result of SLR's and extreme weather events' impact on coastal power infrastructure, such as power stations. Flooding can also impact drinking water supplies and create a habitat for disease-carrying insects like mosquitoes \cite{136}. However, due to the lack of available data about pollution rates pre- and post- extreme weather events, such as storm surges and flooding events, there are not enough studies that model pollution impacts with future SLR scenarios \cite{62}.   

\section{Coastal Assessment}\label{sec: assessment}
\subsection{Coastal Impact Modelling}\label{subsec4: modelling}
The inundation model or bathtub model \cite{16} is a popular and simple model that is used to assess the coastal impacts of SLR and climate-driven events at different spatial scales (i.e.: Global, regional, national, and local). The inundation model is a quantitative model that uses topographic maps to indicate and predict impacted areas by calculating adjacent cells that are under a certain elevation value. Although this model is widely used in research, it is reliant on the accuracy of elevation data and might give false inundation results with low resolution or inaccurate data with high uncertainties. \cite{14} has utilized the inundation model to assess SLR and flooding in coastal areas and highlighted the importance of considering vertical uncertainties inherited in the used digital elevation data. Error correction and transformation methods can also be applied to digital elevation data to increase the reliability and confidence of the analysis results. For example, Aster Global Digital Elevation Map (GDEM), a public source of elevation data has conducted a validation assessment of their data and enhanced the resolution using different horizontal shifts, and elevation error estimation and correction of Band3B error \cite{134}.\\

Variety of researches relied on modeling SLR via inundation. \cite{52} used Light Detection and Ranging (LiDAR) \cite{lidar} remote sensing data to build an inundation model to project different scenarios of SLR in the eastern emirates of UAE. Every 8 pixels in the LiDAR images were grouped to determine the adjacent pixels that should be inundated for each SLR scenario. \cite{110} used an integrated modeling system combining hydrodynamic and hydrological models, to study the impact of SLR and river discharge on estuarine hydrodynamics and ecosystems. The 3D model utilized various tidal and climatic factors with different SLR scenarios. The results suggested that salinization is impacted by the river discharge process more than SLR. However, SLR will have more impact on estuarine hydrodynamics that will eventually alter freshwater transportation into the sea. Al-\cite{61} used the Hydrological model and watershed modeling system HEC-GeoHMS to simulate watersheds from land to sea, which occur due to climatic events, such as rainfall and post-storm floods, as well as other human activities, including aquifers usage. The used data in the simulation included public elevation data, LU/LC from governmental data, and simulated rainfall data measured based on different public repositories.  
SimCLIM \cite{simclim} is an integrated modeling system that is used to assess coastal sensitivity to climate change and variability using biophysical and socioeconomic parameters. This commercialized tool also includes SLR scenario generator that is used to measure the sensitivity of coastal areas to climatic events. This tool is applicable to all spatial scales and various computational requirements depending on the data and resolution \cite{16}. \cite{58} used SimCLIM modeling to project future changes in specific locations. The tool was set to utilize the normalized patterns of Atmosphere-Ocean General Circulation Models (AOGCMs) \cite{aocmg} in a multi-mode by combining GMSL, regional, and local factors \cite{58}. Other ocean modeling tools were used to analyze the impact of SLR in coastal areas, such as HYCOM \cite{hycom} ocean modeling that was developed for the area of the Gulf of Mexico \cite{23}.
Another example of an integrated modeling system that uses biophysical and socioeconomic components is Dynamic Interactive Vulnerability Assessment (DIVA) \cite{58}, which produces quantitative results on the coastal vulnerabilities on spatial scales ranging from regional to local. This model utilizes a number of regional factors that contribute to SLR to measure regional SLR changes. Examples of the adaptation and impact options integrated within DIVA model include flooding, population density, erosion, estimated land for nourishment, wetland loss, and salinization \cite{132}. \cite{67} used biophysical and socioeconomic data from DIVA model with regional variables to estimate the global economic loss and coastal damage under different SLR scenarios. This study estimated a global Gross Domestic Product (GDP) loss of 0.5\% and a 2\% decrease in human well-being under high SLR scenarios. The drawback of this model was the elimination of some of the climate-induced factors that contribute to SLR, such as storm surges effects, and only considering a limited range of adaptation measures \cite{58}. Other models that were used to study the ecological and environmental impacts of SLR and climatic events were the Ecological Landscape Spatial Simulation Models and Sea Level Affecting Marshes Model (SLAMM) \cite{16}.\\

\begin{figure}
\begin{center}
  \includegraphics [width=0.75\textwidth]{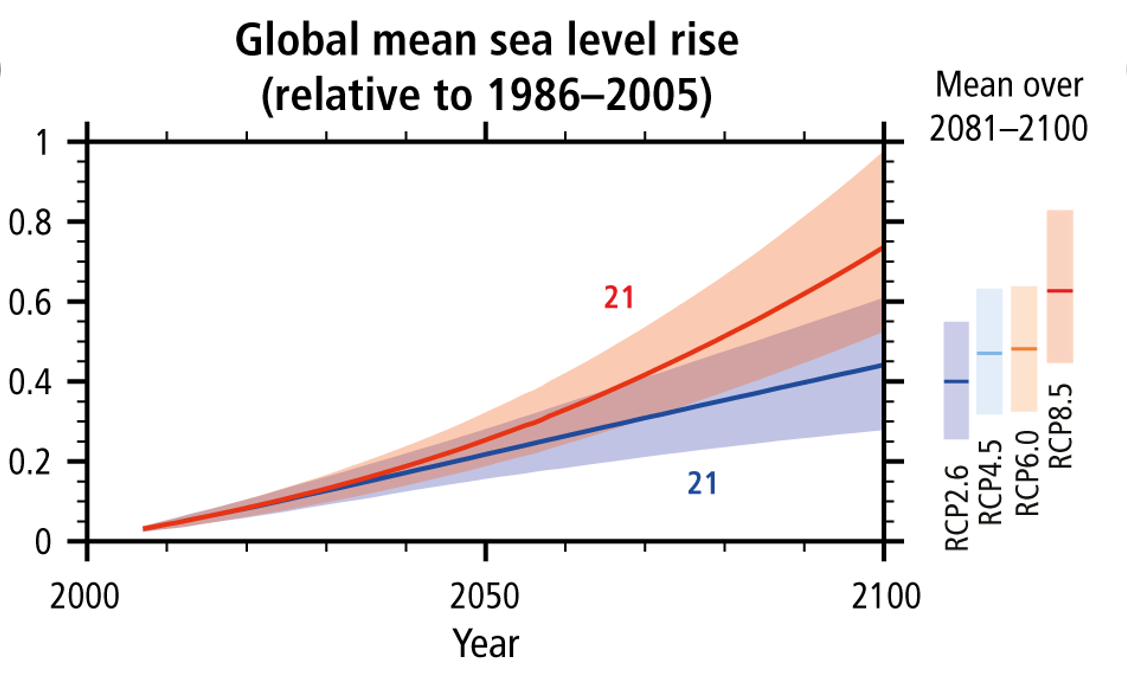}
  \caption{Projection of GMSL (in meters) under different RCP scenarios with associated uncertainties (shading) \cite{IPCC2015}}
  \label{fig: IPCC GMSL scenarios}
\end{center}
\end{figure}

\begin{figure}
\begin{center}
  \includegraphics [width=0.9\textwidth]{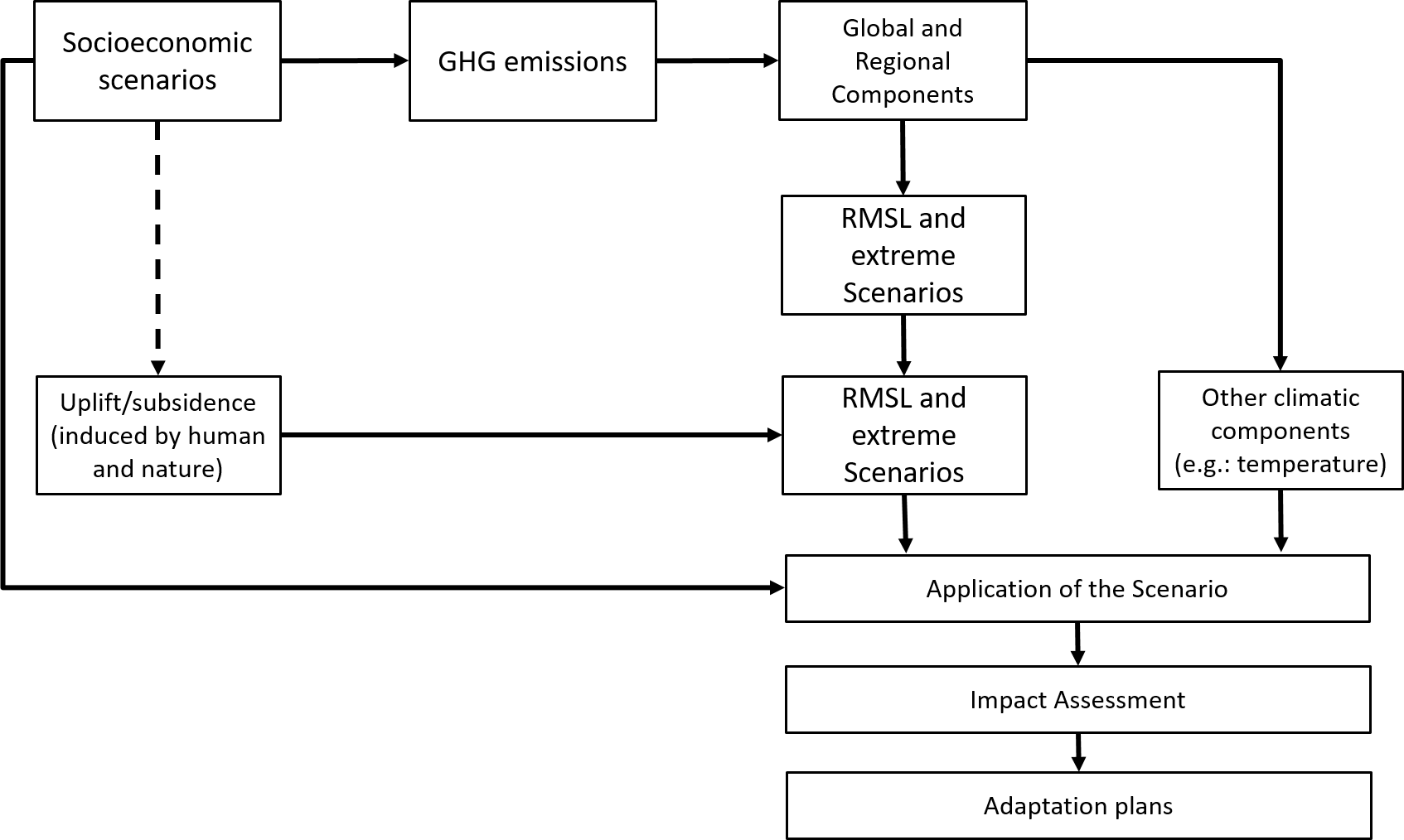}
  \caption{Methodology for developing SLR scenarios for impact, mitigation and adaptation assessments \cite{58}}
  \label{fig: Developing SLR scenario}
\end{center}
\end{figure}

Climate change models make use of future estimations of GHG emissions, as well as ice sheets and glaciers melting to project future SLR based on how sea levels respond to changes in those factors \cite{79}.  Future projections can be based on IPCC's Representative Concentration Pathway (RCP) emission scenarios that estimate the influence of GHG emissions on future global mean and temperature levels. Future projections under RCP2.6., RCP4.5., RCP6.0., and RCP8.5. will increase the global mean temperature by 1, 1.8, 2.2, 3.7 Celsius degrees, and GMSL by 0.4, 0.47, 0.48, 0.63 meters, respectively, by the year 2100 \cite{IPCC2015}. Figure \ref{fig: IPCC GMSL scenarios} shows the projection of GMSL under different RCP scenarios according to IPCC AR5 \cite{IPCC2015}.  \cite{39} used IPCC's RCP emission scenarios, local vertical land movement, and glacio-hydro-isostatic movement to project future flooding scenarios in the Italian peninsula using topographic maps, elevation data, bathymetry data, and Interferometric Synthetic Aperture Radar (InSAR) satellite images. The results obtained from this study showed that coastal areas are highly vulnerable and subject to marine flooding by the year 2100. Vertical land movement data can be extracted from tide gauge data \cite{127}. \cite{21} modeled different scenarios to study the impact of future SLR on the low-lying land and protection structure in the coastline of the Tianjin-Hebei district in China. Three scenarios were modeled according to values obtained from the literature (low, medium, and high) for different years (2030, 2050, and 2100), taking into consideration land subsidence and flooding that can result from extreme storm surge events (occurrence rate of once in 50 years, 100 years, 200 years, and 500 years) which were modeled via Regional Ocean Modelling System (ROMS).
\cite{70} used SLR projection as a part of a systematic planning strategy to evaluate three different adaptation measures that range from hard-engineered protection structures to planning Land Use (LU) and maintenance activities in coastal areas, namely beach nourishment, under extreme scenarios (IPCC RCP8.5.). The results of this projection were showcased to help in decision-making.
\cite{94} surveyed the increasing use of probabilistic models in projecting future SLR scenarios as they consider the uncertainties associated with the contributing factors to regional and local sea level changes, such as mesoscale ocean processes. However, probabilistic projections also have uncertainties due to their dependence on emission scenarios, which are also uncertain.
\cite{47} mentioned that SLR future LU/LC projections were used to evaluate the impacts of storm surges on coastal areas. They also mentioned that statistical and probabilistic Bayesian Networks were used with SLR projections to study the long-term coastal impacts of SLR.
Although \cite{15} did not consider geomorphological and biological impacts of SLR and the exposure of near-shore structures to SLR in their projection scenarios, they admitted the influence of uncertainties associated with climatic and non-climatic events, as they believed that scenarios are built to determine the influence of different factors and not project uncertainties. A number of regional factors were used in SLR projection and visualization due to their huge influence on SLR. With the inclusion of regional components in projection, scenarios can amplify the impacts of global factors, which will be reflected in an increase in the projected SLR levels \cite{25,58}. \cite{25} suggested including factors concerned with the Atlantic Ocean alongside the regional factors to provide comprehensive projections about SLR in the Mediterranean Sea. Figure \ref{fig: Developing SLR scenario} illustrates a standard framework developed by an IPCC working group \cite{58} that can be applied in building SLR scenarios using a set of contributing factors as identified by IPCC.\\

Some studies incorporated AI techniques into SLR projection and simulation. \cite{10} used AI-based visualization of flooding in the province of Quebec-Canada to raise awareness about climate change. Due to the lack of accurate and high-quality flooding data, a simulated environment was created to obtain data in addition to other real data to be used in training a custom Generative Adversarial Networks (GAN). GAN models use a generator to create data based on real data parameters and then use a discriminator to distinguish between the generated data and the real data \cite{gan}. The flooded areas were covered using a painter model to give a realistic flooding scenario \cite{10}. 
\cite{35} re-modeled an extreme SLR scenario that occurred in the Mediterranean Sea 350 AD to visualize how it would impact the current Heraklion port in Crete, Greece. Spatial data that contains topographic, bathymetry and LU data were fed alongside modeled SLR and tsunami waves scenarios into an animation suite to build a 3D visualization tool with scientific output.
Others have used historical data to calculate SLR over the years to analyze the trends causing SLR for future projects. \cite{22} have used Matlab to pre-process and calculate SLR and Mean Sea Level (MSL) trends using linear regression models applied on historical tide gauge data collected from tide stations' sensors in Indonesia \cite{22}. \cite{27} performed error correction on historical tide gauge data of Hong Kong coasts and used multiple regression models to calculate changes in sea levels. \cite{138} conducted a comparative study on the use of historical GMSL data for forecasting using different machine and deep learning algorithms, where the results showed that deep learning algorithms such as Dense Neural Network (DNN) and WaveNet Convolutional Neural Networks provided more reliable results than linear regression.
\subsection{Coastal Vulnerability Assessment}\label{subsec4: CVA}
Part of studying the impacts of SLR on coastal areas is to assess to which extent they are vulnerable to SLR. The term "vulnerability" is used to define the susceptibility level of coastal areas when being exposed to certain climate hazards \cite{74}. This term also encompasses concepts such as risk, resilience, and adaptability \cite{112}. Different analytical approaches can be used to assess the degree of vulnerability by incorporating elements of different aspects, such as physical, socioeconomic, and ecological. Furthermore, vulnerability assessments are spatially scale-dependent, therefore the spatial scale of the study area (regional, national, or local) should be taken into consideration. Vulnerability can be defined as "The degree to which a system, sub-system, or component is likely to experience harm due to exposure to a hazard, either a perturbation or stress" \cite{74}, with it having three components: exposure, sensitivity, and adaptive capacity \cite{74,112}. The exposure (e.g.: population density and LU) and sensitivity (e.g.: inundation depth) components are used to measure the impact of climate hazards while the adaptive capacity component (e.g.: disaster management and response) is used to identify the ability of socioeconomic factors (e.g.: humans, infrastructure, habitats, \dots, etc.) to adapt to changes following climate hazards \cite{112}. \cite{74} have identified indicator-based approaches as the main coastal vulnerability assessment tools in events related to climate change. Index-based vulnerability assessment methods are a sub-category of indicator-based methods that focus on simplifying the complex interaction among different parameters of various aspects and representing them in a simpler and more understandable format, to help decision-makers plan proper responses. Another sub-category of indicator-based methods is variable-based methods, which focus on studying a set of independent variables related to coastal issues. Many researchers relied on using index-based approaches to assess coastal vulnerabilities. Index-based approaches include Coastal Susceptibility/sensitivity Index (CSI), Coastal Vulnerability Index (CVI), Coastal Risk Index (CRI), and socioeconomic Vulnerability Index (SVI) \cite{46}. CVI is the most popular index-type approach in research that is used to study and assess the coastal vulnerabilities of specific study areas \cite{49}. A standardized and quantitative way of assessing CVI was developed by Gornitz \cite{gornitz1991}. His approach uses geometric average (eq. \ref{eq:gornitz}) of different biophysical parameters, including geomorphology, shoreline change rate (i.e. erosion and accretion), coastal slope, regional SLR, mean significant wave height, and mean tidal range \cite{73}.

\hspace{1in}
\begin{equation}\label{eq:gornitz}
CVI=\frac{\sqrt{x_{1} \times x_{2}\times x_{3} \times ... \times x_{n}}}{n}
\end{equation} 
\emph{where:}\\
\emph{$x$ represents a parameter}\\
\emph{$n$ refers to the total number of parameters used in the CVI assessment}\\ 
\hspace{1in}

Gornitz's CVI assessment approach was widely utilized and adopted by others \cite{73}. Thieler and Hammer-Klose (defined the standard U.S. Geological Survey (USGS) CVI assessment) \cite{97}, \cite{98} used similar physical parameters to assess both positive and negative coastal responses to climate change driven events \cite{79}. Coastal assessment methods are applicable to regional, national, and local spatial scales, and can provide a detailed assessment of different segments of the shoreline.\\

Index-based approaches that depends on Gornitz's method used a variety of physical parameters to assess coastal vulnerability with a variation of the selected variables, according to the nature of the study area. On a national level, 
\cite{36} stated that The Department of Irrigation and Drainage in Malaysia used a CVI method in what is called the National Coastal Vulnerability Index Study (NCVI), to assess the vulnerability of coastal areas to SLR and help policymakers in adopting proper adaptation measures. Nevertheless, a CVI study was conducted on Tok Jembal beach using Thelier and Hammar-Klose ranking scheme but with 3 physical variables and climatic-related components. The set of used parameters included geomorphology, coastal slope, regional SLR, temperature, wind speed, and wave's type. Other CVI based works included variables from other non-climatic parameters, such as environmental and socioeconomic \cite{75,81}. \cite{99} (eq.\ref{eq:McLaughlinandcooper}) believed in the importance of including socioeconomic factors in CVI assessment, thus they developed a CVI technique that incorporates parameters related to climatic components, socioeconomic, and physical parameters. \\

\hspace{1in}
\hspace{1in}
\begin{equation}\label{eq:McLaughlinandcooper}
\displaystyle CVI=\frac{CC\ sub\!-index + CF\ sub\!-index + SE\ sub\! -index}{3}
\end{equation} 
\emph{Where:} \\
\emph {CC: Coastal Characteristic sub-index associated with physical/biophysical parameters}\\
\emph {CF: Coastal Forcing sub-index associated with climatic components}\\
\emph {SE: socioeconomic sub-index}\\
\hspace{1in}

\cite {100} used \cite{cooper1998} guidelines to conduct a vulnerability of ecosystem habitats, such as mangrove and seagrass beds in some countries in Africa and the Pacific Islands. The assessment involved 19 different parameters that fall under three categories: exposure, morality, and adaptive capacity. The study found that the regional factors, namely the continuous tectonic uplift movement in the study area of Tanzania led it to have higher resilience to climate change and SLR compared to the other areas involved in the study. Vulnerability assessments were also used to evaluate economic risk and market shock via Economic Vulnerability Assessment (EVA) model \cite{26} that was developed by UN and Fedri using different socioeconomic attributes that fall under either exposure (size, specialization, location) or shocks (trade and natural shocks) \cite{115}. Coastal Sensitivity Index (CSI) was also used in a number of research works to include ecosystem and oceanographic parameters, in addition to the physical parameters used in CVI assessments \cite{56}. \\

Generally, the selection of the type and number of indexes will be reflected in the assessment results, and the selected index set can vary according to the specification of the studied areas. \cite{77} listed some of the criteria that should be considered in choosing an index-based assessment approach, such as the availability of the data associated with the selected index and its robustness to be influenced by other indexes. Further, the number of selected indexes should be sufficient to obtain reliable assessment results \cite{77}. Table \ref{tab:CVI comparison} shows different types of parameters that were used in research. 

\begin{longtable}{R{2cm} c R{1.5cm} l R{5cm}}
\caption{Comparison of Coastal Vulnerability Assessment (CVI) techniques in literature}\label{tab:CVI comparison} \\ \hline
Paper & Year & Region & Spatial Scale & Approach \\ \hline
\endfirsthead
\caption{\textit{(Continued)} Comparison of Coastal Vulnerability Assessment (CVI) techniques in literature}\\ \hline
Paper & Year & Region & scale & Approach \\ \hline
\endhead \hline
\multicolumn{5}{r}{\textit{Continued on next page}} \\
\endfoot \hline
\endlastfoot
\cite{97} & 1999 & US Atlantic coast & National & CVI (Geomorphology, Shoreline change-rate, Coastal-slope, regional SLR, Mean significant wave-height, Mean tidal-range)\\

\cite{99} & 2010 & Northern Ireland & Local & CVI coastal characteristics (resilience and susceptibility)+ coastal forcing + socioeconomic factors \\

\cite{76} & 2015 & Lithuania in the south-eastern Baltic Sea & Local & CVI are combined with DS(the outcome analytical hierarchical process (AHP)) \\

\cite{98} & 2016 & peninsular coastline of Spain & National & CVI (Geomorphology, Shoreline change-rate, Coastal slope, regional SLR, Mean significant wave height, Mean tidal range)\\

\cite{49} & 2018 & Italy  & Local & CVI with 10 parameters (1)Geologic (Geomorphology, Coastal slope, Shoreline Erosion/accretion, Emerged beach width, Dune width), 2) Physical process(regional SLR, Mean significant wave height, Mean tide range), 3) Vegetation(Width of vegetation behind the beach, Posidonia oceanica)) \\

\cite{121} & 2018 & Hawaiian Islands  & Local & CVI and InVEST model to calculate Exposure Index (EI). parameters: bathymetry, shoreline geomorphology, regional SLR, wind and wave actions, LU/LC, population \\

\cite{73} & 2019 & Andhra Pradesh (CAP) region in India & Local & CVI (Geomorphology, Shoreline change-rate, Coastal-slope, regional SLR, Mean significant wave-height, Mean tidal-range)\\

\cite{75} & 2019 & Bangladesh & National & CVI method of Mclaughlin and Cooper(2010) \cite{99} that consists of three sub-indices: 1) coastal characteristics vulnerability sub-index; 2) coastal forcing vulnerability sub-index 3) socioeconomic vulnerability sub-index \\

\cite{69} & 2020 & Sultanate of Oman & National & CVI (Coastal geomorphology
Coastal slope, Coastal elevation, Tidal range, Bathymetry) \\

\cite{36} & 2021 & Malaysia's east coast, Terengganu State beaches & Local & CVI of coastal vulnerabilities using Hammar-Klose and Thieler CVI rankings \\

\cite{111} & 2021 & South India & Regional & CVI with 10 parameters: geomorphology, shoreline erosion/accretion rate, coastal slope, regional SLR, mean significant wave height, mean tide range, storm wave run-up, regional elevation, LU/LC change, mean wave height\\

\cite{112} & 2021 & South Korea & National & CVI of 3 main components: Exposure(population density and age group distribution, coastal industrial facilities, GRDP), Sensitivity (inundated depth and impacted areas), Adaptive capacity (humans, emergency response and disaster management, relief fund, public officials)\\

\cite{141} & 2022 & Nigeria & National & CVI using physical and socioeconomic parameters (geomorphology, coastal slope, bathymetry, wave height, mean tidal range, shoreline change rate, regional SLR, population, cultural heritage, LU/LC, and road network)\\

\cite{140} & 2023 & Northern area of the estuary of Sebou’ in Morocco & Local & CVI with machine learning algorithms (geomorphology, elevation, slope, shoreline change, natural habitat, SLR, maximum wave height, and tidal range)\\

\end{longtable}

\section{Mitigation and Adaptation Measures (MAM)}\label{sec: mam}

This section discusses Mitigation and Adaptation Measures (MAM) that can be implemented in response to SLR induced by climate change. Overall, those responses can be categorized into mitigation measures as a proactive response to SLR, and adaptation measures that act as both proactive and reactive responses. Those measures include man-made/natural structures and non-structural measures that involve policies and sets of best practices to protect coastal areas and communities. Utilizing both structural and non-structural measures will increase the effectiveness of protection and reduce losses \cite{24}. \\

 \subsection{Structural Adaptation Measures}\label{subsec: Structural Mitigation Measures}
Structural mitigation measures are widely implemented across the globe. Various research works were conducted to study both their long- and short- terms effectiveness in protecting coastal areas against SLR \cite{24}. Examples of structural mitigation measures:
\begin{itemize}
    \item Seawall: a protection structure used to protect the coastal landform from SLR impacts \cite{59}.
    \item Breakwater: it is used to protect coastal zone areas and beach material against strong wave actions. It also enables extending the beach area for different human/economic activities. Breakwaters are built either tangible or parallel to the beach according to the beach's nature, to provide maximum protection. The elevation of the breakwater should be determined according to the height of the local maximum tidal wave \cite{59}. 
    \item Seagrass beds: eco-system-based protection measure of high ecological value that adds means of support to other structural mitigation measures against SLR impacts \cite{24,100,114}.
    \item Dike: structures built along the coast to protect against high wave actions and flooding \cite{24}
    \item Dunes: Piles of sand and/or other materials that are formed due to either natural causes such as wind actions or built constructions (filled with artificial material) to reduce the impact of wave actions and coastal erosion \cite{53}. 
    \item Beach Nourishment: a structural measure that is implemented by filling up the impacted areas with artificial material to protect the coastal areas from SLR impacts. Beach nourishment is costly and may cause a negative impact on the environment, therefore it is usually implemented after conducting assessments studies and strategical evaluation of the affected areas \cite{24}.
\end{itemize}

\cite{48} studied the impact of structural infrastructure at Damietta Promontory - Egypt in protecting coastal areas against SLR using satellite images and historical data on changes in the shoreline. It was found that seawalls and detached breakwater bodies have effectively protected the coastline against SLR. However, it was found that anthropogenic interventions were causing changes to shorelines far greater than SLR. Further, the importance of the maintenance of those structural bodies to maintain the current elevation was also highlighted. Additionally, some recommendations about extending the length of the breaker were provided. \cite{20} also mentioned some of the other existing adaptation measures that are being used to protect against high waves and SLR impacts in the northeastern coast of Egypt, specifically the city of Alexandria. This includes the Muhammed Ali Seawall and the international coastal road that was constructed with a high elevation ($3$m) by the government. The impact of the type of material used to construct the structural protection measures was also discussed in the literature as it determines their efficiency in stopping SLR. \cite{53} used Unmanned Armed Vehicles (UAV) \cite{uav} to study the prior and post outcomes of constructing a dune at Cardiff State Beach - California to protect against flooding and extreme water events. The constructed dune consisted of different materials, including vegetation, sand, cobble, and rip-rap. Although this dune was affected by a storm during the construction process, the post-construction imagery results obtained from UAV during an observation period of 9 months, showed that it had, overall, effectively reduced flooding impacts. In China, a seawall was constructed to protect the coastline of Tianjin-Hebei District against SLR and extreme weather events that occur once every 100 years; however, \cite{21} suggested that the current elevation of this seawall will not be able to survive against such an extreme event by 2030, based on a presented study of future projections of SLR. This study simulated different SLR scenarios including the existence and absence of extreme weather events of different intensities using the Regional Ocean Modelling System (ROMS) and other data related to that area. Seawalls were built in the coastal areas of Mombasa in Kenya by the tourism sectors to protect against the SLR, extreme weather event impacts, and coastal erosion, in addition to maintaining tourism activities \cite{131}. \cite{59} studied the effectiveness of some of the implemented structural adaptation measures at some of UAE's vulnerable coastal areas to prevent erosion that results from SLR and wave activities. Detached breakwaters and offshore barriers were found to be better at protecting the coastal areas against coastal erosion compared to seawalls, as they impacted the beach and were degraded over time due to tidal wave actions. \\

Sediment movement due to Wave actions during storm surges in the Croatian Krk island in the Adriatic Sea led to the formation of beach winter profile that acted as a natural protection structure for the coastal areas against erosion. It was found that this naturally formed protection structure acted better in protecting against coastal erosion than the existing landslide protection wall which was constructed with low quality and is susceptible to collapse under the impact of continuous intense wave actions \cite{54}. Existing wetlands and lakes near the coastal zone in Alexandria-Egypt were considered as a defense line to hold seawater from propagating further into the land in case of the occurrence of an extreme weather event \cite{20}. \cite{114} studied the impact of using different scenarios by combining adaptation measures with natural ecosystems to protect coastal areas. This study relied on using data-based approaches from literature and qualitative-based approaches carried out via field studies in the area Mecklenburg-Germany in the southeast coast of the Baltic Sea by groups of experts. This research found that combining structural adaptation measures with submerged vegetation will result in both positive economic and environmental impact by adding an appealing look and value to the coastal areas, which could attract both tourists and environmental scientists.\\

\subsection{Non-structural Adaptation Measures}\label{subsec: Non-structural Mitigation Measures}
The negative impact of climate change and SLR on islands can be more severe in comparison to coastal states \cite{IPCC96,26}. For example, SIDS, will be highly impacted by environmental changes and SLR \cite{26}. As per IPCC, economically disadvantaged SIDS countries are highly vulnerable to climate change and SLR impacts \cite{IPCC96}; hence, which arises the necessity of implementing proper and effective adaptation measures. Conversely, financially well-off members of SIDS like the Kingdom of Bahrain can adopt non-structural adaptation measures in the form of policies and regulations on energy-related sectors to reduce CO2 and GHG emissions \cite{26}. Due to challenges associated with SIDS,  \cite{18} suggested that those countries use Eco-based Adaptation (EbA) measures which "provide a combination of protect and advance benefits based on the sustainable management, conservation, and restoration of ecosystems" \cite{18}. EbA provides protection by deflecting waves and reducing erosion, thus increasing land elevation via sediment built-ups, which has proven its effectiveness in mitigating SLR impacts \cite{18, 62}. \cite{30} made use of reinforcement learning to build a model that helps policymakers evaluate the economic cost of investing in adaptation infrastructure to SLR. The used model was based on Marcov Decision Process (MDP) model. It provided a proactive measurement by evaluating the cost of implementing a protection infrastructure versus the economic loss as a consequence of SLR impact. The model also considered different SLR scenarios according to NOAA model, the impact of extreme natural events associated with SLR (such as storms and hurricanes), and residents' and corporations' willingness to support the government's decision to implement the SLR adaptation measures. \cite{60} conducted a national survey in the US to measure how local communities perceive different SLR adaptation measures adopted by their local authorities. Fu's definition of adaptation measures was used to create four categories, as follows: 1) Protection 2) Accommodation 3) Managed Retreats 4) Planning. The survey results showed that adaptation measures at coastal localities focus more on planning, which includes hazard plans, coastal and emergency management plans, SLR vulnerability assessments, raising public awareness and education, forming SLR taskforce, and conducting vulnerability assessments. Nevertheless, planning might give a false sense of safety, thus the public preferred having protection measures implemented as part of adaptation measures to SLR. \cite{70} used a participatory approach that involved different stakeholders to evaluate the impact of implementing different adaptation measures in Portugal's coasts to protect them from erosion. Adaptation measures were classified into adaption, nourishment, management, relocation, protection, and restoration. The overall benefit-to-cost ratio was in favor of scenarios that involved protection structures and beach nourishment when considering environmental, economic, and social impacts. \cite{113} gave some adaptation suggestions to be used alongside the existing structural protection measure in Po River Delta - Italy to reduce the impact of SLR, such as farmers switching crops that will yield higher economic returns. \cite{80} evaluated the adaptation measure in Southeast Asian countries, like Malaysia, and identified the existing flaws, including the unavailability of documentation, the absence of a centralized assessment and strategies platform, the gap between assessment and application, and the emphasis on the importance of socioeconomic component in assessments, which is reflected negatively on the produced set of policies and strategies. \\

\begin{table}[t]
\caption{Comparison of different Mitigation and Adaptation Measures (MAM) in literature}
\label{tab:mam comparison}
\begin{tabular}{ p{0.6cm} R{2.6cm} R{5cm} R{2.2cm} }
\hline
\multicolumn{1}{l}{MAM Category}   & Type (\# papers)             & Papers                       & Examples                                      \\ \hline
\multirow{3}{*}{Structural}        & Hard Structures (6)          & \cite{21,54,59,70,20,48,113} & Breakwaters, seawalls, dikes                  \\
                                   & Soft Structures (2)          & \cite{53,54}                 & Dunes, beach nourishment                      \\
                                   & Eco-Based Structures (4)     & \cite{114,18,20,62}          & Seagrass bed, wetlands \& reef conversion and restoration \\ \hline
\multicolumn{1}{l}{Non-Structural} & Policies and Regulations (6) & \cite{26,18,30,41,60,113}    & Relocation , hazard mapping, public awareness \\ \hline
\end{tabular}
\end{table}

Generally, hard-engineered structures are preferred as protection measures to reduce the impact of SLR, wave actions, and extreme weather events; however, they might result in negative socioeconomic and ecological impacts. The built-in protection structures might limit the usage of the coastal areas, which are usually utilized for habitation and touristic activities. This can be even more challenging in cities that are highly dependent on coastal activities and lack the necessary land capacity for relocation, such as SIDS countries \cite{53}. Additionally, some hard structures may reduce the attractiveness of the beach and impact fisheries activities, leading to negative economic impact \cite{59}. A combination of eco-based systems and other structural protection measures can be effective in protecting against SLR physical impacts while maintaining the socioeconomic and ecological value of the coastal areas; however, we find that such scenarios might not be feasible in all regions, due to many local factors like environmental conditions \cite{18}, which play a role in determining the effectiveness of implementing such measures. Some ecosystems, that protect the shoreline against change,s have the ability to adapt to changes in their environment and can cope with SLR up to a certain extent \cite{62}. However, we still question their ability to continue adapting to the recent SLR trends and provide the same level of protection to shorelines against changes. Although stakeholders are aware of the importance of using protection structures in mitigating coastal changes induced by climate change and SLR, they still prefer soft adaptation measures, such as beach nourishment, as they are easier, simpler, and cost-efficient to implement \cite{70}. Overall, the trade-off between economic cost loss, such as the value of land and properties, and implementation costs drives the decision to implement proper adaptation measures to reduce the risk caused by climatic-driven events \cite{132}. At the local scale, communities are also playing a role in protecting their areas by taking proactive measures and supporting local authorities \cite{18,60}. Table \Ref{tab:mam comparison} compares different mitigation and adaptation measures covered in literature over the past 5 years. furthermore,

\section{Sea Level Changes in the Arabian Gulf}\label{sec: Arabian Gulf}
The Arabian Gulf is a semi-enclosed sea located in Western Asia, extending from the Gulf of Oman and the Indian Ocean through the Straight of Hormuz. Its bathymetry ranges from 5 meters for shallow zones to 94 meters for the deepest zones. It is surrounded by eight countries: United Arab Emirates (UAE), Qatar, Kingdom of Saudia Arabia (KSA), Kingdom of Bahrain, Sultanate of Oman, Kuwait, Iran, and Iraq. The majority of those countries have prosperous economies driven by their substantial oil reserves \cite{17}. The coastal population and activities in these countries have also witnessed substantial growth, leading to an increasing demand for coastal urbanization \cite{59}. 
The increasing SST in the Arabian Gulf due to a number of climatic and non-climatic drivers resulted in significant impacts in the region \cite{145}. In addition to its environmental impact including coral bleaching, extension of fish species, and salinity changes, the increasing evaporation rate, driven by the increase in sea and air temperatures, resulted in the physical and chemical processes that altered the intensity and frequency of climatic events occurrence. Trend analysis showed that there are homogeneous changes in SST throughout the Arabian Gulf with visible seasonal patterns in summer and autumn seasons, specifically in the inner northwestern zone of the Gulf that varies from the outer zone near the Hormuz strait \cite{146}. An unobserved extreme SST of 37.8$\textsuperscript{o}$ C was recorded near Kuwait's coasts in 2020 \cite{62}. The difference in water levels between the Arabian Gulf and the Gulf of Oman, driven by high evaporation rates, which results in a movement of a water mass of 416 km$\textsuperscript{3}$/y through the Strait of Hormuz into the Arabian Gulf \cite{17, bruciaferri2022gulf18}. Difference in water pressure also plays a significant role in inward and outward water movement through the Strait of Hormuz \cite{sealevelvariability}. 
Additionally, the growing population and demand for water supply, coupled with limited freshwater resources, have led to an increase in the number and operations of seawater desalination plants along the coastal areas of the Arabian countries in the Gulf. This has resulted in an increasing salinity of seawater in beach areas and contributed to rising steric sea levels \cite{al2017climate}. Furthermore, wind actions and extreme events caused by Shamal wind during the winter season can influence local tidal wave patterns, potentially resulting in a sea-level rise of a few meters \cite{chow2022combining} with evaporation being higher in the winter season compared to the summer despite the high temperature during this season \cite{sealevelvariability}. In summer, Kaus wind, in addition to wind draft influenced by variations of temperature between land and water, causes complex variability in short-term sea level changes and drives surface pollutants towards the coastal areas. The freshwater budget of the Arabian Gulf is driven by evaporation, precipitation, and freshwater discharge, particularly in the northwestern region \cite{sealevelvariability}. The flow of water through the Strait of Hormuz occurs in two layers: the upper layer consists of an inward flow from the Gulf of Oman and the Indian Ocean toward the Arabian Gulf, including seasonal surface variations. The second layer (bottom layer) consists of an outward flow toward the Gulf of Oman and includes saltier water \cite{campos2020freshwater}. non-eustatic sea level rise can be triggered by the occurrence of sudden events such as earthquakes generated from the north and east zones or tsunamis generated by extreme events in the Arabian Sea and the Indian Ocean \cite{56}. The timeline of changes in sea levels in the Arabian Gulf can be traced back to the late Quaternary when the Gulf experienced a reflooding, reaching its peak of more than 1 meter above current sea levels during the mid-Holocene. Sea levels gradually decreased to reach their present levels by late-Holocene \cite{historicalsealevel}. \\

The majority of the northeastern shorelines of the Arabian Gulf near the Hormuz Strait are protected by the Hajar and Zagros mountains, while the shorelines within the Gulf basin contain various landforms, including sandy beaches, mudflats, mangroves, seagrass beds, and sabkhas \cite{vaughan2019arabian}. The geomorphology of these areas serves as a natural barrier against sea level rise (SLR) in some regions, while in other areas it makes them vulnerable to variations in sea levels. In low-lying areas, offshore barriers, including both natural and man-made structures, have been employed to reduce wave velocity and intensity \cite{59}.
Satellite images data and modeling tools enabled the monitoring of changes and impacts of sea levels in the Arabian Gulf. An example of this is the GULF18-4.0 Model which simulates various oceanographic features with a high resolution of 1.8 km and optimizes the vertical grid for important physical processes, including SST, waves, and tides \cite{bruciaferri2022gulf18}. 
A combination of hydrodynamics and spectral wave models was customized to incorporate regional factors such as the wind/wave effects induced by the Shamal wind, the characteristics of the inner water channel regions, and the submerged areas of the Arabian Gulf, with a focus on studying the southeastern region \cite{chow2022combining}. Satellite altimetry were used to study  changes in sea level SST \cite{17, 145, sealevelvariability}

\section{Discussion and Conclusion}\label{sec: conc}
There is no doubt that climate change is a major force that is leading to overall negative changes worldwide. The rise in temperature increases the melting rates of ice sheets and glaciers. It also causes a temperature rise of ocean water that propagates both horizontally and vertically, which impacts both the underwater environment and coastal areas. One of the perceived changes to those global temperature changes is SLR. Further, climate change leads to changes in other climatic events, such as the increase in the occurrence and intensity of extreme weather events, like storm surges and rainfall rates. Additionally, these climatic changes directly influence wave and wind actions, leading to rapid changes in shorelines and increasing the coastal areas' exposure to inundation and flooding events. Nevertheless, studies showed that non-climatic regional effects can have a noticeable influence on SLR and climatic events \cite{7} (e.g.: tsunamis are triggered by an earthquake \cite{54,62}). Those components include vertical land movement, land-based ice melting, water exchange between land and sea, and elastic response to mass loss of land ice melting. Although the percentage of contribution of the different climatic and non-climatic components to SLR can vary, they should be considered in future projections and scenario-building topology to obtain realistic results.\\ 

It is worth mentioning that uncertainty associated with those components should be considered for future projections. Even though the earlier IPCC reports did not consider them in future projections, recent studies incorporated them in estimating future SLR change and its related components. Generally, the majority of studies use IPCC RCP emission scenarios in estimating future GMSL and customize them to calculate regional SLR trends according to the specifications of the studied areas. Other studies used scenarios developed by NOAA and other research groups (e.g.: NCA) or relied on historical regional data to measure the regional SLR. Studies have also found a deviation between RMSL and GMSL rates when incorporating regional factors and uncertainties. This is to be explained by the fact that SLR are not uniform globally and they depend on the geographical location. Furthermore, the propagation of changes in GMSL are timescale-dependent. For example, the contribution of Greenland's ice melting to SLR will be observed in North America before other regions. The relation between the geographical location and propagation of changes is perhaps what led \cite{116} to identify some of the regional factors, according to findings of studies, as global contributing factors. 
Other climatic Regional factors, such as ENSO \cite{65} can play a significant role in SLR both globally and regionally. Nevertheless, we believe that SLR projections and scenarios should incorporate both regional factors and uncertainties associated with global factors that contribute to GMSL to provide a comprehensive analysis and reliable estimations of the impacts of SLR. Those uncertainties and deviations in estimated results should be presented clearly by researchers without any conflict of interest to avoid the consequences of false results. Therefore, researchers should present their analysis outcomes to policymakers as a proactive measure in mitigating events driven by climate change. Additionally, visualization tools should be utilized to showcase the impacts of climate change and SLR, towards appropriate planning and decision-making, as well as awareness campaigning. \\

Data requirements vary according to the purpose of the study and the number and type of parameters to be used in the analysis. Spatial data (Satellite images and digital elevation data) were used in the majority of SLR studies as they can be easily obtained from a number of public repositories, such as AsterGDEM and USGS. Using other types of data adds more value to the assessment SLR impacts. Examples of other data that can be used for assessments are population, LU/LC, shoreline geomorphology, and climatic and non-climatic factors. However, the research works on this topic suffer from the lack of accurate and recent data needed for projection and visualization. Simulation tools were used to cope with the lack of available data associated with different contributing factors, while visualization tools were used to interpret changes and impacts of climate-driven events. Several studies used simulation tools to generate data related to contributing factors, such as wave action and rainfall, to be used in impact assessment (see section \ref{sec: factors}). Historical data were used to make future projections about GMSL and RMSL to assess their impacts and propose adaptation measures. Those data were fed into different models and tools for processing. machine learning algorithms such as linear regression were also used in SLR forecasting studies. However, a recent comparative study by \cite{138} proved that deep learning provides more reliable results in forecasting GMSL based on historical data. Although the use of deep learning algorithms is plausible, neither machine learning nor deep learning studies did not consider the uncertainties associated with the contributing factors to SLR. SLR trends have accelerated in the twentieth century \cite{116}, which raises the question of to which extent the historical data is reliable in obtaining realistic results. \\

The impacts of SLR and its associated climatic events on physical, socioeconomic, 
 and environmental/ecological/chemical aspects are correlated and perceived negatively except for the ecological aspects, in which some ecosystems were impacted positively with the increase of SLR and SST. The level of SLR impact relies on a number of characteristics of the studied area such as the geographical location and geomorphology of the coast which determines the type of regional factors that influence SLR and the coast's vulnerability to it. Socioeconomic components also help in determining the level of impact and can be a driving force toward implementing MAM policies. According to \cite{IPCC96}, Southeast Asian countries are classified as highly vulnerable to SLR, as they suffer from increasing frequency and intensity of extreme weather events and have vulnerable communities. We found that the majority of the papers reviewed in this study were about the Mediterranean Basin and Southeast Asian countries which match with the distribution of vulnerable areas and communities to SLR as shown in figure \ref{fig: impacted areas worldwide}. Nevertheless, it is worth mentioning that SLR is not the only source of physical changes on shoreline as anthropogenic interventions impacts can exceed those caused by SLR and climatic events. \\
 
Index-based approaches were used widely in assessing coastal vulnerability to SLR impacts while coastal modeling approaches were used for assessing SLR impacts. Each assessment approach used a different set of contributing factors and other parameters to assess coastal areas based on the spatial scale, available data, objective of the assessment, and processing requirements. The results of those assessments are used to suggest the proper MAM strategies to be implemented. It was noticed that the studies conducted on the Mediterranean Basin countries and the U.S. area rely more on adaption structures, in comparison to those of Southeast Asian countries, where adaptation policies are mainly seen as their MAM strategy. Nevertheless, we believe that both MAM strategies should be implemented to reduce potential damage. An example of long-term planning of MAM strategies was found in the UAE \cite{42}, where several protection structures were built at the vulnerable segments of coasts and national monitoring platforms are used to monitor and forecast climate hazards. Additionally, adaptation plans and policies were placed as part of the national development plan toward achieving sustainability. 
Some final remarks to be considered for future researches: 
\begin{itemize}
    \item Uncertainties associated with global contributing factors to be properly Incorporated into GMSL and RMSL projections and updated versions of sea level budget to be used (as per data availability)
    \item GMSL is not uniformly distributed globally; hence time and location dimensions should be considered when estimating individual contributing factors for SLR projections.
    \item Geomorphological changes on coastal zones due to anthropogenic interventions to be isolated from SLR impacts during coastal assessment studies as changes caused by prior can surpass the latter leading to false results and interpretation of the current status.
    \item The lack of available data is limiting studies on SLR impacts on ecosystems and chemical aspects (e.g.:  groundwater contamination, seawater pollution after flooding and natural disasters). 
    \item There is a limit to the extent of considering historical data reliable in forecasting future SLR, where significant changes of trends in climatic events and GMSL are evident (noting that historical data are not sufficient on their own in forecasting future SLR and other factors and their associated uncertainties must be considered )
\end{itemize}

\bibliographystyle{cas-model2-names}

\bibliography{main}



\end{document}